\documentstyle[preprint,eqsecnum,aps]{revtex}
\begin{document}
\draft
\title{THE EQUATION OF STATE FOR COOL RELATIVISTIC\\
TWO-CONSTITUENT SUPERFLUID DYNAMICS.}
\author{Brandon Carter and David Langlois}
\address{D\'epartement d'Astrophysique Relativiste
et de Cosmologie, C.N.R.S.,\\
Observatoire de Paris, 92195 Meudon, France.}
\date{\today}
\maketitle
\begin{abstract}
The natural relativistic generalisation of Landau's two constituent superfluid
theory can be formulated in terms of a Lagrangian ${\cal L}$ that is given as
a function of the entropy current 4-vector $s^\rho$ and the gradient
$\nabla\!_\rho\,\varphi$ of the superfluid phase scalar.
It is shown that in the ``cool" regime, for which the entropy is attributable
just to phonons (not rotons), the Lagrangian function ${\cal L}\{\vec s,
\nabla\!\varphi\}$ is given by an expression of
the form ${\cal L}=P-3\psi$ where $P$ represents the pressure
as a function just of $\nabla\!_\rho\,\varphi$ in the (isotropic) cold
limit. The entropy current dependent contribution $\psi$ represents
the generalised pressure of the (non-isotropic) phonon gas, which is
obtained as the negative of the corresponding grand potential energy
per unit volume, whose explicit form has a simple
algebraic dependence on the sound or ``phonon" speed
$c_{_{\rm P}}$ that is determined by
the cold pressure function $P$.
\end{abstract}

\section{Introduction}
\label{sec:1}

The purpose of the present work is to derive the natural {\it cool limit
form} of the equation of state that is needed to complete the
formulation of the natural relativistic generalisation of Landau's two
constituent superfluid theory \cite{[1]}.  The qualification ``cool" is
to be understood here as referring to the low temperature limit in which
account is taken only of lowest order deviations from a nearby {\it
cold} configuration in which thermal effects are absent altogether.

Whereas the cold (strictly zero temperature) superfluid case is
described by irrotational configurations of a simple perfect fluid model
(i.e.  one that is both isotropic and barytropic), on the other hand the
allowance for thermal effects, even in the cool limit, necessitates the
use of a two constituent fluid model, of the kind that was pioneered by
workers such as London, and perfected by Landau.  Accurate treatment of
non-stationary applications would require allowance for viscosity of the
``normal" constituent.  However it is sufficient for many purposes --
and entails no loss of accuracy at all in the case of equilibrium states
such as simple cylindrical vortex configurations -- to use a strictly
conservative treatment as in Landau's original model \cite{[1]}.

In addition to the neglect of dissipation, the main simplification on which
the analysis below is based is the neglect of the non-linear excitations
known as ``rotons" which are of dominant importance in the ``warm" regime
nearer to the phase transition (beyond which lies the regime of ``hot"
states in which superfluidity is absent altogether). The presence of such
effects makes it very difficult to derive the equation of state over the
full range of the three relevant variables, which in addition to the
temperature $\Theta$ say, could be taken to consist of the relevant
conserved particle number density $n$ say, and the relative velocity
${v}$ of the two constituents.  Although it is out of the question for
neutron star matter, an experimental investigation of the equation of
state should be feasible in practice for the laboratory case of
Helium-4.  However even in this experimentally accessible case, the
exploration of the relevant phase space (as parametrised by $\Theta$,
$n$, and ${v}$) does not yet appear to have been sufficiently systematic
and thorough (see e.g. \cite{[1b]}).

In contrast with the dubious quality of present day knowledge of ``warm"
superfluid dynamics,  the state of knowledge of the ``cool" limit is very
satisfactory. In this limit, complications such as ``rotons" can be
ignored, the only important thermal effects being entirely attributable to
simple ``phonon" excitations, which can be adequately described by an
essentially linear treatment whose original development is again largely
attributable to the (by now experimentally well substantiated) work of
Landau \cite{[2]}. The detailed development of the necessary statistical
mechanics has been summarised in a convenient form by Khalatnikov
\cite{[3]}.  It will be shown below that although it was originally
carried out in a non-relativistic framework, the nature of this
statistical analysis is such that it can be translated into a
relativistic form without any change of form provided that sufficient
care is used in defining the appropriate variables for the relativistic
version.  In consequence, as shown in the appendix, Landau's expressions
for the first and second sound speed can also be retained without change
in the relativistic regime.

\section{Relativistic generalisation of 2-constituent superfluid dynamics}
\label{sec:2}

It has recently been shown[4] that the natural relativistic generalisation
of the standard Landau theory of two constituent superfluid dynamics in its
simplest strictly conservative version[1] can be formulated in a
particularly convenient manner by taking as the starting point a
Lagrangian scalar ${\cal L}$ that is given as an appropriate function
(which was denoted by $X$ in the original presentation \cite{[4]}) of
the entropy current vector $s^\mu$ and of the gradient
\begin{equation}
   \mu_\rho= \hbar\nabla\!_\rho\,\varphi \label{eq:2.1}
\end{equation}
of the superfluid phase scalar $\varphi$.  Two of the required equations
of motion are obtained in the form of the usual conservation laws as
given by
\begin{equation}
 \nabla\!_\rho n^\rho=0  \ , \label{eq:2.2}
\end{equation}
for the particle current $n^\rho$ constructed below, and
\begin{equation}
  \nabla\!_\rho s^\rho=0 \ .\label{eq:2.3}
\end{equation}
The system is completed by the equation
\begin{equation}
 s^\rho\nabla\!_{[\rho}\Theta_{\sigma]}=0 \label{eq:2.4}
\end{equation}
(using square brackets to denote index antisymmetrisation) governing the
evolution of a certain {\it thermal momentum} covector $\Theta_\rho$. This
last equation can alternatively be expressed (in a form that is useful as a
starting point for the derivation of corresponding circulation and helicity
conservation laws \cite{[5]}) as the vanishing of the Lie derivative of the
thermal momentum vector with respect to the associated {\it temperature}
vector, $\beta^\rho$, i.e.
\begin{equation}
  \beta^\sigma\nabla\!_\sigma\Theta_\rho+\Theta_\sigma\nabla\!_\rho
   \beta^\sigma =0\ ,\label{eq:2.5}
\end{equation}
where
\begin{equation}
  \beta^\rho= -\big(s^\sigma\Theta_\sigma\big)^{-1}s^\rho\ .\label{eq:2.6}
\end{equation}
The thermal 4-momentum covector $\Theta_\rho$ and the particle number
current vector $n^\rho$ are specified in this formulation as the algebraic
functions of the entropy current vector $s^\rho$ and of the superfluid
4-momentum covector $\mu_\rho$ that are obtained by partial differentiation
of the Lagrangian according to the infinitesimal variation formula
\begin{equation}
   d{\cal L}=\Theta_\rho ds^\rho-n^\rho d\mu_\rho \ . \label{eq:2.7}
\end{equation}

\section{Alternative action functions and the stress momentum energy tensor}
\label{sec:3}

The formulation that has just been summarised has the technical advantage of
being particularly economical in so much as it involves only 5 independent
component variables, namely the phase scalar $\varphi$ and the 4 independent
components of the entropy current vector $s^\rho$. This feature of economy has
recently been exploited for the purpose of setting up a correspondingly
economical Hamiltonian formulation of the theory \cite{[6]}, and it has also
been exploited as a guide to the formulation of an analogously economical
theory for describing thermal effects in superconducting cosmic strings
\cite{[7]}.

Although mathematically equivalent as far as its physical (``on shell")
solutions are concerned, this more recent 5-component formulation differs
from the earlier convective variational formulation \cite{[8],[9]} (which
has the alternative advantage of being less specialised) that treated the
current vectors $n^\rho$ and $s^\rho$ on the same footing as independent
variables, thus involving a total of 8 independent space time components.
The presence of the three extra components made it necessary to include,
as an extra dynamical equation, the irrotationality condition
\begin{equation}
  \nabla\!_{[\rho}\mu_{\sigma]}=0 \label{eq:3.1}
\end{equation}
that is the Poincar\'e integrability condition for the existence of a
potential $\varphi$ satisfying (\ref{eq:2.1}). In the convective variational
formulation $\mu_\rho$ is on the same footing as $\Theta_\rho$, both
being obtained by partial differentiation according to the formula
\begin{equation}
   d\Lambda=\Theta_\rho d s^\rho+\mu_\rho d n^\rho ,\label{eq:3.2}
\end{equation}
where the convective variational master function $\Lambda$ is obtainable
from the Lagrangian ${\cal L}$ of the newer more specifically adapted
approach by a Legendre type transformation given by the relation
\begin{equation}
 \Lambda = {\cal L}+n^\rho\mu_\rho . \label{eq:3.3}
\end{equation}

The approach based on the Lagrangian ${\cal L}$ that will be used here is in
fact a compromise, intermediate between the convective variational approach
based on the master function $\Lambda$, and another independently developped
(Clebsch type) variational approach \cite{[10],[11]} in which both currents
are treated as dependent variables, their specification being given by a
generalised pressure function $\Psi$ in terms of $\mu_\rho$, as defined by
(\ref{eq:2.1}) and of $\Theta_\rho$, which in this version has independent
status, by the partial differentiation formula
\begin{equation}
 d\Psi=-s^\rho d\Theta_\rho-n^\rho d\mu_\rho \label{eq:3.4}
\end{equation}
where the generalised pressure function $\Psi$ itself is given by
\begin{equation}
  \Psi={\cal L}-\Theta_\rho s^\rho\ . \label{eq:3.5}
\end{equation}

Although the generalised pressure function $\Psi$ will not act in a
fundamental role in the formulation used here, it nevertheless plays an
important part, notably in the expression for the stress momentum energy
tensor.  An important part is also played by a set of three scalar
functions of state, ${\cal C}$, ${\cal B}$, and ${\cal A}$ (respectively
describable as the {\it caloric} coefficient, the {\it bulk} coefficient,
and the {\it anomaly} coefficient), that are defined as matrix elements
in the relation expressing the momenta in terms of the velocities in the
form
\begin{equation}
  \Theta^\rho={\cal C}s^\rho+{\cal A}n^\rho \ , \qquad\quad
\mu^\rho={\cal A}s^\rho+{\cal B}n^\rho .\label{eq:3.6}
\end{equation}
These coefficients can be used to convert the canonical expression
\begin{equation}
   T^\rho_{\ \sigma}=n^\rho\mu_\sigma+s^\rho\Theta_\sigma+\Psi g^\rho_{\
   \sigma} \label{eq:3.7}
\end{equation}
for the stress momentum energy density tensor of the 2-constituent
superfluid into the manifestly symmetric though not so elegant form
\begin{equation}
 T^{\rho\sigma}={\cal B}n^\rho n^\sigma +{\cal A}\big( n^{\rho} s^{\sigma}
 +s^{\rho} n^{\sigma}\big)+{\cal C}s^\rho s^\rho +\Psi g^{\rho\sigma}\ ,
  \label{eq:3.8}
\end{equation}
where $g^{\rho\sigma}$ is the contravariant inverse of the -- flat or
curved -- space time metric tensor $g_{\rho\sigma}$ that is to be used
for index raising or lowering.  For the purpose of the present work it
will be more convenient to work with an alternative recombination
expressible in terms of the {\it dilation} coefficient $\Phi^2$ and the
{\it determinant} coefficient ${\cal K}$ by
\begin{equation}
  T^{\rho\sigma}=\Phi^2\big(\mu^\rho \mu^\sigma+{\cal K}s^\rho s^\sigma
\big)+\Psi g^{\rho\sigma} \ , \label{eq:3.9}
\end{equation}
where
\begin{equation}
 \Phi^2={\cal B}^{-1}\ ,\hskip 0.8 cm {\cal K}={\cal C}{\cal B}-{\cal A}^2
  \ .\label{eq:3.10}
\end{equation}

\section{The cold limit and the speed of sound}
\label{sec:4}

Before considering the treatment of the lowest order thermal corrections,
it is to be recalled that in the cold (i.e. strictly zero temperature)
limit, the superfluid behaves as a perfect fluid of simple barytropic type.
This means that in terms of the unit flow vector $u^\rho$ defined by
\begin{equation}
  n^\rho= nu^\rho\ ,\qquad\quad  u^\rho u_\rho=-c^2 \ ,\label{eq:4.1}
\end{equation}
where $c$ is the speed of light, the stress tensor reduces to the isotropic
form
\begin{equation}
  T^{\rho\sigma}=\big(\rho+c^{-2}P\big)u^\rho u^\sigma+P g^{\rho\sigma}\  ,
  \label{eq:4.2}
\end{equation}
in which the mass-density $\rho$ and the (isotropic) pressure $P$ are
directly related by a single variable equation of state. The equation of
state can be specified either by giving the rest frame energy density
$\rho$ as a function of the corresponding particle number density $n$ in
the form
\begin{equation}
  \rho=\rho\{n\} \ , \label{eq:4.3}
\end{equation}
or equivalently, in dual form, by giving the pressure $P$ as a function of
the corresponding effective mass (i.e. relativistic chemical potential)
variable $\mu$ in the form
\begin{equation}
  P=P\{\mu\} \ , \label{eq:4.4}
\end{equation}
(using small curly brackets to indicate functional dependence, as distinct
from the multiplication  that would be indicated by ordinary brackets).
The connection between these two formulations is given by the familiar
differential specifications
\begin{equation}
 \mu={d\rho\over dn}\ ,\qquad n={1\over c^2}{dP\over d\mu}\ ,\label{eq:4.5}
\end{equation}
and the dually symmetric relation
\begin{equation}
  \rho+c^{-2}P\ =n\mu  \ . \label{eq:4.6}
\end{equation}

For the simple (``barytropic") perfect fluid model that has just been
formulated, the convective variational master function $\Lambda$ referred to
in the introduction is given directly by the first version of the equation
of state as
\begin{equation}
  \Lambda=-c^2\rho\{n\} \, \label{eq:4.7}
\end{equation}
while the corresponding dual Lagrangian ${\cal L}$, which in this zero
temperature limit has just the same form as the Clebsch type potential
function $\Psi$, will be  given analogously by the dual version of the
equation of state as
\begin{equation}
  {\cal L}= \Psi=P\{\mu\} \ . \label{eq:4.8}
\end{equation}
To complete the translation into the notation of the preceeding section
the particle momentum covector will be given simply by
\begin{equation}
  \mu_\rho=\mu u_\rho\ , \label{eq:4.9}
\end{equation}
and of course the entropy vector $s^\rho$ and the thermal momentum covector
$\Theta_\rho$ will both vanish, while the temperature vector $\beta^\rho$ is
singular in this zero temperature limit.

One of the most important quantities in this simple perfect fluid model is
the characteristic sound or ``phonon" speed, $c_{_{\rm P}}$ say, i.e. the
short wavelength limit of the propagation velocity relative to the preferred
rest frame of small (necessarily longitudinal) perturbations.  This
characteristic speed is immediately derivable from either of the versions
(\ref{eq:4.3}) or (\ref{eq:4.4}) of the equation of state via the
familiar differential formula
\begin{equation}
  c_{_{\rm P}}^2= {dP\over d\rho}= c^2{n\over \mu}{d\mu\over dn} \ .
  \label{eq:4.10}
\end{equation}
This can be used to construct the phonic (or sonic) metric tensor
\begin{equation}
  {\cal G}^{\rho\sigma}=g^{\rho\sigma}+\Big({1\over c^2}
     -{1\over c_{_{\rm P}}^2} \Big)u^\rho u^\sigma\ , \label{eq:4.11}
\end{equation}
whose null eigen covectors $p_\rho$, as defined by
\begin{equation}
  {\cal G}^{\rho\sigma}p_\rho p_\sigma=0\ ,\label{eq:4.12}
\end{equation}
are tangential to the characteristic hypersurfaces of sound propagation
in the medium, in the same way as the null covectors of the ordinary
spacetime metric $g^{\rho\sigma}$ are tangential to are the
characteristic hypersurfaces of light propagation in vacuum.

\section{The concept of ``normal" and ``superfluid" density contributions}
\label{sec:5}

As soon as we want to allow for deviations from the cold limit case that
has just been summarised, it is necessary to increase the number of
independent variables in the fundamental equation of state from 1 to 3.
It might naively have been hoped that 2 would suffice for the cool limit
with which we are concerned here but, as will be made apparent at once,
this ceases to be possible unless one is willing to restrict attention to
states for which there is no relative motion between the two constituents.
In the convective variational formulation based on $\Lambda$ as the
fundamental state function, the 3 independent variables correspond to the 3
scalar invariants that can be constructed from the pair of independent
current vectors, namely $n^\rho$ whose direction determines the particle
or ``Eckart" rest frame, and $s^\rho$ which similarly determines the
thermal or ``normal" rest frame, so that an obviously convenient way of
choosing the 3 independent variables in a fundamental state function of
the form
\begin{equation}
  \Lambda=\Lambda\{n,x,s\} \label{eq:5.1}
\end{equation}
will be to take them to consist of the particle rest frame number
density $n$ as given by
\begin{equation}
  c^2 n^2=-n_\rho n^\rho \ ,\label{eq:5.2}
\end{equation}
the cross product variable $x$ as given by
\begin{equation}
  c^2 x^2=-n_\rho s^\rho \ ,\label{eq:5.3}
\end{equation}
and the thermal rest frame entropy density $s$ as given by
\begin{equation}
  c^2 s^2=-s_\rho s^\rho \ .\label{eq:5.4}
\end{equation}

In the dual formulation based on the lagrangian ${\cal L}$ that will be used
here, the relevant independent variables are the 3 scalar invariants that
can be constructed from the entropy current vector $s^\rho$  and the
momentum covector $\mu_\rho$ whose orientation determines what is known in
this context as the ``superfluid" rest frame.  It is obviously convenient
to take one of the required scalars to be again the ``normal" entropy
density variable $s$ as defined by (\ref{eq:5.4}), while taking the
other two to consist of a new cross product variable $y$ given by
\begin{equation}
   c^2 y^2 =-\mu_\rho s^\rho \ ,\label{eq:5.5}
\end{equation}
together with the effective mass variable $\mu$ given by
\begin{equation}
   c^2\mu^2=-\mu_\rho\mu^\rho \ ,\label{eq:5.6}
\end{equation}
so that the required generalisation of (\ref{eq:4.8}), i.e. the relevant
analogue of (\ref{eq:5.1}), will be given by an expression of the form
\begin{equation}
  {\cal L}={\cal L}\{\mu,y,s\} . \label{eq:5.7}
\end{equation}

Starting from an expression of this form it can be seen from
(\ref{eq:3.6}) and (\ref{eq:3.10}) that the secondary variables $n^\rho$
and $\Theta_\rho$ will be given in terms of the primary variables
$\mu_\rho$ and $s^\rho$ of this formulation by
\begin{equation}
  n^\rho=\Phi^2\big(\mu^\rho -{\cal A} s^\rho\big)\  ,\qquad
   \Theta_\rho=\Phi^2\big( {\cal K}s_\rho + {\cal A} \mu_\rho\big) \ .
   \label{eq:5.8}
\end{equation}
where by (\ref{eq:2.7}) the relevant dilation, determinant and anomaly
coefficients $\Phi^2$, ${\cal K}$, and ${\cal A}$ are given by the
partial differentiation formulae
\begin{equation}
 c^2\Phi^2={1\over\mu}{\partial {\cal L}\over\partial\mu}\ ,\qquad
 c^2\Phi^2{\cal K}=-{1\over s}{\partial{\cal L}\over\partial s}\ ,\qquad
  c^2\Phi^2{\cal A}=-{1\over 2 y}{\partial{\cal L}\over\partial y}
  \ .  \label{eq:5.9}
\end{equation}

It was a generic Lagrangian function of this form (\ref{eq:5.7}) that
was the starting point for the construction, in implicit form, of a
Hamiltonian reformulation of the theory \cite{[6]}.  However in order to
be able to make such a construction explicit it is necessary to sacrifice
generality and deal separately with particular kinds of state function.
{}From a purely mathematical point of view the simplest kind that can be
envisaged is the``regular" category, as characterised \cite{[9]} by the
condition that the anomaly parameter ${\cal A}$ vanishes, which
evidently, by the original definition (\ref{eq:3.6}), means that the
momenta are aligned with the corresponding current vectors.  It can be
seen from (\ref{eq:5.9}) that in the formulation based on the Lagrangian
${\cal L}$ that we are using here this regularity condition is
expressible as the condition that the state function should depend only
on $\mu$ and $s$ but not on the third cross product variable $y$.
However although it can be hoped that such a mathematically attractive
alignment ansatz may provide a useful approximation in contexts
involving hot conducting fluids, it is certainly not at all an
appropriate simplification in the context of superfluidity.

In order to see how to obtain a more suitable form of equation of state
it will be useful to establish an appropriate translation relating the
terminology of the traditional formalism that was developed in a
non-relativistic framework to that of the relativistic formalism used here.
In particular it will be convenient to introduce appropriate relativistic
generalisations of the concepts of the ``normal" and ``superfluid" mass
densities ${\rho_{_{\rm N}}}$ and ${\rho_{_{\rm S}}}$. One way of
defining such a generalisation is to base it on a corresponding vectorial
decomposition \cite{[4]} of the particle current $n^\nu$, which in the
non-relativistic limit \cite{[12]} is interpretable as being proportional
to the Newtonian mass current $\rho^\nu=m n^\nu$ for some constant
coefficient $m$ representing the mass per particle.  However such a
definition has the disadvantage that in a relativistic context there will
be a certain margin of ambiguity in the choice of the appropriate ``rest
mass" parameter $m$.  It will be more convenient for most purposes,
including that of the present work, to adopt a natural alternative
possibility that gives the same result in the non-relativistic limit but
that is free of any such ambiguity of normalisation.

The definition that will be adopted here -- and that seems clearly most
appropriate in the relativistic context -- is based on the decomposition
of the stress momentum energy density tensor in the form
\begin{equation}
 T^{\rho\sigma}={{\rho_{_{\rm S}}}\over{\mu_{_{\rm
   N}}}^2}\mu^\rho\mu^\sigma+ {{\rho_{_{\rm N}}}\over {s_{_{\rm
    S}}}^2}s^\rho s^\sigma+\Psi g^{\rho\sigma}\ ,\label{eq:5.10}
\end{equation}
where ${s_{_{\rm S}}}$ is the value of the entropy density in the
``superfluid" rest frame defined by $\mu^\rho$, and ${\mu_{_{\rm N}}}$
is the analogously defined superfluid chemical potential with respect to
the ``normal" rest frame defined by $s^\rho$, i.e.
\begin{equation}
  {s_{_{\rm S}}}=s\Big(1-{{v}^2\over c^2}\Big)^{-1/2}={y^2\over\mu}\ ,
\qquad {\mu_{_{\rm N}}}=\mu\Big(1-{{v}^2\over c^2}\Big)^{-1/2}={y^2\over s}
 \ , \label{eq:5.11}
\end{equation}
in which the velocity ${v}$ is the relative translation speed between the
``normal" and ``superfluid" frames as given by
\begin{equation}
   {{v}^2\over c^2}=1-{s^2\mu^2\over y^4}\ .  \label{eq:5.12}
\end{equation}

The definitions implicit in (\ref{eq:5.10}) are such as to ensure that
the momentum density ($T{^{\,^0}_{\ ^3}}$ say) in the direction of the
relative motion will have magnitude ${\rho_{_{\rm N}}} {v}$ in the
``superfluid" frame, while it will have magnitude $\rho_{_{\rm S}} v$ in
the in the ``normal" frame.  (One could define a corresponding
generalisation of the Newtonian concept of the mass current by setting
$\rho^\nu=$ ${\rho_{_{\rm S}}}{\mu_{_{\rm N}}}^{-1}\mu^\nu+ {\rho_{_{\rm
N}}}{s_{_{\rm S}}}^{-1}s^\nu$, but it would be of limited utility since
it would not retain the exact conservation property of its Newtonian
limit.) It can be seen by comparison with (\ref{eq:3.9}) that the values
of the quantities so defined can be evaluated in terms of the partial
derivatives appearing in (\ref{eq:5.9}) using the formulae
\begin{equation}
  {\rho_{_{\rm S}}}=\Phi^2{\mu_{_{\rm N}}}^2\ ,
 \qquad {\rho_{_{\rm N}}}=\Phi^2{\cal K}{s_{_{\rm S}}}^2 \ .\label{eq:5.13}
\end{equation}

\section{Statistical mechanics of the phonon gas}
\label{sec:6}

As in the standard Landau analysis \cite{[2]}, it will be supposed that,
in the cool limit with which we are concerned, the thermal excitations
can be described as a bosonic gas of non interacting phonons.

To see how this works out, and to fix the notation, we start by
recalling that the esssential step in such an analysis \cite{[2]} is the
evaluation of the expected value of the occupation numbers $\nu_r$, say
where $r$ is an index running over the relevant quantum states.  The
standard trick for doing this is work in terms of groups, with
collective index $R\{r\}$ say, consisting of a large number, $G_{_R}$
say, of nearby states characterised by approximately equal quantum
numbers, the point of this being to ensure that although the mean
occupation number $\bar\nu_{_R}$ may be small, the total group
occupation number,
\begin{equation}
  {\cal N}_{_R}=G_{_R}\bar\nu_{_R}=\sum_{r:R\{r\}=R} \nu_r \label{eq:6.1}
\end{equation}
will be large compared with unity. This makes it possible to replace the
exact Bose statistical formula
\begin{equation}
  \exp\{S\}=\prod_R{(G_{_R}+{\cal N}_{_R}-1)!\over(G_{_R}-1)!\,
  {\cal N}_{_R}!}\ , \label{eq:6.2}
\end{equation}
for the total entropy $S$ of the whole system (in units such that
Boltzmann's constant is set to unity) by the corresponding Stirling formula
\begin{equation}
  S=\sum_R G_{_R}\Big(\big(\bar\nu_{_R}+1\big)\ln\big\{\bar\nu_{_R}+1\big\}
    -\bar\nu_{_R}\ln\bar\nu_{_R}\Big) \ .\label{eq:6.3}
\end{equation}
Introducing the dimensionless state functions
$\sigma\{r\}=\sigma\big\{R\{r\}\big\}$ defined by setting
\begin{equation}
 \bar\nu_{_R}={1\over {\rm e}^{\sigma\{R\}} -1}\ , \qquad \sigma\{R\}=
   \ln\Big\{ {\bar\nu_{_R}+1\over\bar\nu_{_R}}\Big\} \ ,\label{eq:6.4}
\end{equation}
the formula (6.3) can be usefully rewritten as
\begin{equation}
 S=\ln {\cal Z}+\sum_R {\cal N}_{_R}\sigma\{R\} \ , \label{eq:6.5}
\end{equation}
where the zustandssumme, ${\cal Z}$, is defined as a sum over all
admissible combinations $\{\nu_r\}$ of individual occupation numbers by
\begin{equation}
  {\cal Z}=\sum_{\{\nu_r\}}\exp\Big\{\sum_r\nu_r\,\sigma\{r\}\Big\}
     \ ,\label{eq:6.6}
\end{equation}
from which the logarithm required in (\ref{eq:6.5}) is obtained simply as
\begin{equation}
  \ln {\cal Z}=\sum_R G_{_R}\ln\{\bar\nu_{_R}+1\} .\label{eq:6.7}
\end{equation}
The corresponding variation formulae are thus obtained even more simply
as
\begin{equation}
  \delta\ln{\cal Z}=-\sum_r \nu_r\delta\sigma\{r\}\ , \hskip 0.8 cm
\delta S=\sum_r\sigma\{r\}\,\delta\nu_r \ . \label{eq:6.8}
\end{equation}

Having carried out these routine preliminary steps, one can immediately
obtain the thermal equilibrium state specified by maximising the entropy
subject to possible constraints on the total occupation number $N$ and
the total energy momentum covector ${\cal P}_\rho$ as defined in terms
of the corresponding microscopic energy momentum covectors $p_{\,\rho}$
by
\begin{equation}
   {\cal N}=\sum_R {\cal N}_{_R}=\sum_r\nu_r\ , \qquad
   {\cal P}_\rho=\sum_r \nu_r\, p_\rho\{r\}\ .\label{eq:6.9}
\end{equation}
In order to satisfy the ensuing requirement
\begin{equation}
  \delta S+\alpha\delta {\cal N} +\beta^\rho\delta{\cal P}_\rho= 0\ ,
  \label{eq:6.10}
\end{equation}
where the scalar $\alpha$ and the 4 vectorial components $\beta^\rho$
are Lagrange multipliers, it evidently follows from (\ref{eq:6.8}) that the
solution for the equilibrium distribution must be given simply by
\begin{equation}
  \sigma=-\alpha-\beta^\rho p_\rho \ ,\label{eq:6.11}
\end{equation}
which by (\ref{eq:6.5}) implies the thermodynamic relation
\begin{equation}
 S-\ln{\cal Z}=-\alpha{\cal N}-\beta^\rho{\cal P}_\rho \ .\label{eq:6.12}
\end{equation}

In the case of phonons, as in that of the more familiar example of Planckian
photons, there is no conservation law imposing any constraint on the total
occupation number so the multiplier $\alpha$ will vanish, while the vector
$\beta^\rho$ will give the temperature $\Theta$ and relative flow velocity
components $v^i$ ($i$=1,2,3) with respect to the chosen reference frame
according to the specifications
\begin{equation}
  \alpha=0\ ,\qquad \beta^{_0}={1\over\Theta}\ , \qquad
    \beta^i={1\over\Theta} v^i\ .\label{eq:6.13}
\end{equation}

For the actual evaluation of the distribution thus obtained it is of
course convenient to work in the continuum limit for which the summation
goes over to a phase space integration in the simple form
\begin{equation}
  \sum_R G_{_R}=\sum r\rightarrow {\cal V}\int{ dp_{_1}dp_{_2}dp_{_3}
    \over (2\pi\hbar)^3}\ ,\qquad
   {\cal V}=\int dx^{_1}dx^{_2}dx^{_3} \ .\label{eq:6.14}
\end{equation}
This is the first point at which the phonon case with which we are concerned
here deviates from the more familiar photon gas, in which an extra factor of
2 would be required to allow for the existence of 2 (transverse)
polarisation modes, whereas in the phonon case there is only a single
(longitudinal) polarisation mode.

The only integration for which (\ref{eq:6.14}) is actually needed in
practice is that for the logarithm of the zustandssumme as given by
(\ref{eq:6.7}), since once this is known the other thermodynamic
quantities involved will be obtainable from it by straightforward
partial differentiation procedures.  Defining the {\it grand potential}
energy $\mit\Omega$, the {\it free} energy $F$, and the ordinary energy
$U$ of the system by formulae of the standard form \cite{[2]}
\begin{equation}
  {\mit\Omega}=-\Theta\,\ln{\cal Z}\ , \qquad F=U-\Theta S\ ,
\qquad U=-{\cal P}_{_0} \ ,\label{eq:6.15}
\end{equation}
(which differs however from the terminology of Khalatnikov \cite{[3]}
who uses the symbol $F$ in place of ${\mit\Omega}$) the relation (6.12)
can be rewritten as
\begin{equation}
  F-{\mit\Omega}=v^i{\cal P}_i \ ,\label{eq:6.16}
\end{equation}
while the variation laws (\ref{eq:6.8}) (which are of course to be understood
as determined with respect to fixed boundary conditions, and hence
in particular for a constant value of the volume ${\cal V}$)
give rise to corresponding variation formulae of the form
\begin{equation}
  \delta {\mit\Omega}=-S\delta\Theta-{\cal P}_i\,\delta v^i \ ,\qquad
  \delta F= -S\delta\Theta+v^i\delta{\cal P}_i\ ,\qquad
  \delta U=\Theta\delta S+v^i\delta{\cal P}_i \ .\label{eq:6.17}
\end{equation}

Although it is simpler in so far as polarisation is concerned, on the other
hand the phonon case is more complicated than that of photons in that the
presence of the background medium implies a breakdown of Lorentz invariance
which is expressed by the need to use not the ordinary metric but the
phonic metric (\ref{eq:4.11}) in the relevant nullity restiction
(\ref{eq:4.12}).  In order to be able to satisfy this condition by
expressing the phonon energy $\epsilon$ by the simple formula
\begin{equation}
  \epsilon=-p_{\,_0} = c_{_{\rm P}} p\ ,\qquad p^2=
     p^{\, 2}_{_1}+p^{\, 2}_{_2}+p^{\, 2}_{_3} \label{eq:6.18}
\end{equation}
where $c_{_{\rm P}}$ is the zero temperature sound speed as given by
(\ref{eq:4.10}), we must now restrict the choice of reference frame to
be one in which the underlying superfluid is at rest.  This contrasts
with the case of a photon gas, which is subject to an analogous formula,
with $c$ replacing $c_{_{\rm P}}$, in an arbitrarily boosted frame.
Further restricting the frame by choosing space axes aligned with the
relative flow direction, one obtains the logarithm of the zustandssumme,
in the explicit form
\begin{equation}
 \ln {\cal Z}=- {\cal V} \int{2\pi p^2\sin\theta\, d\theta\,
   dp\over(2\pi\hbar)^3} \ln\big\{1-e^{-\sigma} \big\} \ ,\label{eq:6.19}
\end{equation}
with
\begin{equation}
  \sigma=-\beta^\rho p_\rho={c_{_{\rm P}} p-v p_{_3}\over\Theta}\ ,
     \label{eq:6.20}
\end{equation}
where
\begin{equation}
 v^{_1}=v^{_2}= 0\ ,\qquad v^{_3}=v\ ,\qquad p_{_3}=p\cos\theta\ .
   \label{eq:6.21}
\end{equation}

\section{Thermodynamics of the phonon gas}
\label{sec:7}

To relate the thermodynamic quantities obtained in the preceeding section
to the superfluid continuum variables described in the earlier sections,
we start by introducing the generalised pressure function $\psi$
determined by the grand potential ${\mit\Omega}$ of the phonon gas,
together with the entropy density ${s_{_{\rm S}}}$ in the superfluid
frame --~which is of course to be identified with the component given by
(\ref{eq:5.11})~-- according to the specifications
\begin{equation}
  \Omega=-{\cal V}\psi\ ,\qquad S={\cal V}{s_{_{\rm S}}} \,\label{eq:7.1}
\end{equation}
while the
corresponding preferred frame components ${\cal E}$ and ${\mit\Pi}$ of the
phonon energy and momentum density are specified by setting
\begin{equation}
 {\cal P}_{_0}=-{\cal V}{\cal E}\ , \qquad {\cal P}_{_3}={\cal V}
   {\mit\Pi}\ , \qquad {\cal P}_{_2}={\cal P}_{_1}=0\ .\label{eq:7.2}
\end{equation}
With these definitions the global thermodynamic equilibrium condition
(\ref{eq:6.12}) can be rewritten in terms of the corresponding local
field variables as
\begin{equation}
  {\cal E}+\psi=\Theta\,{s_{_{\rm S}}}+{\mit\Pi}\,v \ ,\label{eq:7.3}
\end{equation}
while by (\ref{eq:6.17}) the fundamental entropy maximisation condition
(\ref{eq:6.10}) gives the relevant local version of the first law of
thermodynamics in the form
\begin{equation}
  \delta{\cal E}=\Theta\,\delta{s_{_{\rm S}}}+v\,\delta{\mit\Pi} \ ,
   \label{eq:7.4}
\end{equation}
whose conjugate, by (\ref{eq:7.3}), is evidently
\begin{equation}
  \delta\psi={s_{_{\rm S}}}\,\delta\Theta+{\mit\Pi}\,\delta v \ .
   \label{eq:7.5}
\end{equation}

Using this last formula one can obtain the values of ${s_{_{\rm S}}}$
and ${\mit\Pi}$ as functions of $\Theta$ and $v$ by partial
differentiation of the generalised pressure function $\psi$, which is
obtained from (\ref{eq:6.19}) in the form
\begin{equation}
 \psi=\Big({3\over\tilde\hbar c_{_{\rm P}}}\Big)^3
  \Big({\Theta\over 4}\Big)^4 \big(1-{v^2\over c_{_{\rm P}}^2}\big)^{-2}
  \label{eq:7.6}
\end{equation}
where $\tilde\hbar$ is a constant that is given, within a numerical factor
that is extremely close to unity, by the usual Dirac Planck constant
$\hbar$, its exact expression being
\begin{equation}
  \tilde\hbar=\Big({1215\over 128\pi^2}\Big)^{1/3}\hbar= {9\over 4\pi}
\Big({5\pi\over 6}\Big)^{1/3}\hbar\simeq 0.99\hbar\ .\label{eq:7.7}
\end{equation}
One thus obtains the expressions
\begin{equation}
   {s_{_{\rm S}}}=\Big({3\Theta\over 4\tilde\hbar c_{_{\rm P}}}\Big)^3
    \big(1-{v^2\over c_{_{\rm P}}^2}\big)^{-2} \label{eq:7.8}
\end{equation}
and
\begin{equation}
{\mit\Pi}=\Big({3\over 4\tilde \hbar}\Big)^3\Big({\Theta\over c_{_{\rm
  P}}}\Big)^4 \big(1-{v^2\over c_{_{\rm P}}^2}\big)^{-3} {v\over c_{_{\rm
   P}}} \ ,\label{eq:7.9}
\end{equation}
which are formally identical with the analogous expressions
as originally derived in a non-relativistic framework \cite{[3]}.

It is to be remarked that the temperature vector (\ref{eq:6.13}) can be
used to rewrite (\ref{eq:7.6}) in the neater form
\begin{equation}
\psi=\Big({3\over 4\tilde\hbar}\Big)^3c_{_{\rm P}}\big(2{\cal G}^{-1}_{\
   \rho\sigma} \beta^\rho\beta^\sigma\big)^{-2}\ ,\label{eq:7.10}
\end{equation}
where ${\cal G}^{-1}_{\ \rho\sigma}$ is the covariant inverse of the
sonic metric given by (\ref{eq:4.11}).

\section{Conclusion: the equation of state for cool superfluid dynamics}
\label{sec:8}

We can use the expressions obtained in the previous section to evaluate
the ``normal'' mass density contribution ${\rho_{_{\rm N}}}$ as introduced
in (\ref{eq:5.10}) in a manner that is formally consistent with the
traditional definition \cite{[4]} by identifying the momentum density
${\mit\Pi}$ with the corresponding mixed component of the stress
momentum energy density tensor in the preferred superfluid frame defined
by $\mu_\rho$ according to the specification
\begin{equation}
  T{^{_0}_{\ _3}}={\mit\Pi}={\rho_{_{\rm N}}} v \ .\label{eq:8.1}
\end{equation}
This leads to a result that is expressible in the form
\begin{equation}
  {\rho_{_{\rm N}}}={4\tilde\hbar\over 3c_{_{\rm P}}}{s_{_{\rm S}}}^{4/3}
    \big(1-{v^2\over c_{_{\rm P}}^2}\big)^{-1/3} \ .\label{eq:8.2}
\end{equation}

We are now in a position to obtain the equation of state for the
required cool superfluid Lagrangian ${\cal L}$ by integrating the
partial differential equation
\begin{equation}
  {1\over s}{\partial{\cal L}\over\partial s}=
     -{c^2\over{s_{_{\rm S}}}^2}{\rho_{_{\rm N}}} \label{eq:8.3}
\end{equation}
obtained from (\ref{eq:5.9}) and (\ref{eq:5.13}).  This leads to the
principal result of this work which is the formula
\begin{equation}
  {\cal L}=P\{\mu\}-3{\psi}\{\mu,y,s\}\ ,\label{eq:8.4}
\end{equation}
where ${\psi}$ is the generalised pressure function of the phonon
gas, which can be seen from (\ref{eq:7.10}) to be given by
\begin{equation}
  3{\psi}=\tilde\hbar c_{_{\rm P}}^{-1/3}\vert {\cal G}^{-1}_{\ \rho\sigma}
   s^\rho s^\sigma\vert^{2/3} \ .\label{eq:8.5}
\end{equation}
Since the operation $\partial/\partial s$ is defined in terms of the
scalar variables $\mu$, $y$, and $s$, the verification that
(\ref{eq:8.4}) satisfies (\ref{eq:8.3}) requires the conversion of
(\ref{eq:8.5}) into terms of these three scalars.  This can be done
using the explicit expression
\begin{equation}
{\cal G}^{-1}_{\ \rho\sigma}=g_{\rho\sigma}+{1\over c^2}\Big(1-{c_{_{\rm
P}}^2\over c^2}\Big) u_\rho u_\sigma\ ,\label{eq:8.6}
\end{equation}
for the covariant version of the phonic metric (\ref{eq:4.11}), which gives
\begin{equation}
{\cal G}^{-1}_{\ \rho\sigma}s^\rho s^\sigma =\big(c^2-c_{_{\rm P}}^2\big)
  \Big({y^2\over\mu}\Big)^2-c^2 s^2 \ .\label{eq:8.7}
\end{equation}
The solution of (\ref{eq:8.3}) is of course not unique (since one can always
add in any function of $\mu$ and $y$) but (\ref{eq:8.4}) is the only solution
compatible with the boundary requirements that it should go over correctly
to the cold limit value (\ref{eq:4.8}) when both $s$ and $y$ vanish, and
that the energy in the static limit $v=0$ should differ from that of the
cold solution with the same particle number density $n$ by an amount
given by the phonon gas contribution ${\cal E}$, which in the static
limit is the same as $3{\psi}$.

We conclude by remarking that the description of $\psi$ as the generalized
pressure function of the phonon gas is justified by the fact that  the
basic cool equation of state formula (\ref{eq:8.4}) translates into
terms of the total generalised pressure function $\Psi$, as defined by
(\ref{eq:3.5}), simply as
\begin{equation}
  \Psi=P+{\psi}\ ,\label{eq:8.8}
\end{equation}
where the appropriate expression for the thermal correction term
$\psi$ is
\begin{equation}
  {\psi}={c_{_{\rm P}}\over 4}\Big({3\over 4\tilde\hbar}\Big)^3\big(
{\cal G}^{\rho\sigma}\Theta_\rho\Theta_\sigma\big)^2 \ ,\label{eq:8.9}
\end{equation}
in which it is to be recalled that (as described in Section 4) $c_{_{\rm
P}}$ and ${\cal G}^{\rho\sigma}$ are given as functions (only) of the
superfluid momentum covector $\mu_\rho$ in a manner determined by the
original equation of state for the pressure $P$ in the zero temperature
limit.  The corresponding expression for the relationship between the
entropy current vector $s^\rho$, the temperature vector $\beta^\rho$,
and the thermal momentum covector $\Theta_\rho$ is
\begin{equation}
  s^\rho=4{\psi}\,\beta^\rho=2\Big({3\over 4\tilde\hbar}\Big)^{3/2}
\big(c_{_{\rm P}}{\psi}\big)^{1/2}{\cal G}^{\rho\sigma}\Theta_\sigma
\ .\label{eq:8.10}
\end{equation}
The first of these relations can be used to verify that the temperature
vector $\beta^\rho$ introduced as a Lagrange multiplier in Section 6
agrees with the earlier definition (\ref{eq:2.6}).

\acknowledgments

The authors wish to thank G.L.~Comer and I.M.~Khalatnikov for
stimulating conversations.

\appendix
\section*{Evaluation of the first and second sound speeds}

To obtain the characteristic speeds of propagation of small perturbations,
it suffices to follow the lines already developed in the context of more
general multiconstituent fluid theory \cite{[9]} using a technique
originally due to Hadamard. The speeds in question are those characterising
the set of possible directions for the normal covector, $\lambda_\rho$ say,
of a characteristic hypersurface of the system. The Hadamard analysis
postulates that the relevant independent dynamical variables, in this
case $s^\rho$ and $\mu_\rho$, should be continuous across the
characteristic hypersurface, but that their derivatives have
infinitesimal discontinuities that will be expressible in the form
\begin{equation}
 \big[\nabla\!_\rho s^\sigma\big]=\lambda_\rho\hat s^\sigma\ ,\qquad
\big[\nabla\!_\rho\mu_\sigma\big]=\lambda_\rho\hat\mu_\sigma\ ,\label{eq:A1}
\end{equation}
in terms of a corresponding set of infinitesimal discontinuity amplitudes
$\hat s^\rho$ and $\hat\mu^\rho $.  For the dependent variables $n^\rho$
and $\Theta_\rho$, one then obtains the analogous relations
\begin{equation}
 \big[\nabla\!_\rho n^\sigma\big]=\lambda_\rho\hat n^\sigma\ ,\qquad
 \big[\nabla\!_\rho\Theta_\sigma\big]=\lambda_\rho\hat\Theta_\sigma \ .
 \label{eq:A2}
\end{equation}

For a generic Lagrangian of the form (\ref{eq:5.7}) the independent dynamical
variables $\mu_\rho$ and $s^\rho$ will determine the dependent dynamical
variables $n^\rho$ and $\Theta_\rho$ by the relation (\ref{eq:5.8}), which
can be rewritten succinctly as
\begin{equation}
  n^\rho=B\mu^\rho-As^\rho \ ,\qquad
\Theta^\rho=Cs^\rho + A \mu^\rho , \ ,\label{eq:A3}
\end{equation}
with
\begin{equation}
   B=2{\partial {\cal L}\over c^2\partial \mu^2}\ , \qquad
   C=-2{\partial {\cal L}\over c^2\partial s^2}\ ,   \qquad
   A=-{\partial {\cal L}\over c^2\partial y^2}\ . \label{eq:A4}
\end{equation}
It follows that the independent discontinuity amplitudes
$\hat\mu_\rho$ and $\hat s^\rho$ will determine the corresponding
dependent discontinuity amplitudes $\hat n^\rho$ and $\hat\Theta_\rho$
by an analogous linear relation of the form
\begin{equation}
 \hat n_\rho= B_{\rho\sigma}\hat\mu^\sigma\ -A_{\sigma\rho}\hat s^\sigma,
  \qquad \hat\Theta_\rho=C_{\rho\sigma}\hat s^\sigma +
   A_{\rho\sigma}\hat\mu^\sigma\ , \label{eq:A5}
\end{equation}
with
\[  B_{\rho\sigma}=Bg_{\rho\sigma}
  -2{\partial B\over c^2\partial \mu^2} \mu_{\rho} \mu_{\sigma}
  +4{\partial A\over c^2\partial \mu^2} \mu_{(\rho} s_{\sigma)}
  +{\partial A\over c^2\partial y^2} s_{\rho} s_{\sigma}\ ,  \]
\begin{equation}
  C_{\rho\sigma}=Cg_{\rho\sigma}
  -2{\partial C\over c^2\partial s^2} s_{\rho} s_{\sigma}
  -4{\partial A\over c^2\partial s^2} s_{(\rho} \mu_{\sigma)}
  - {\partial A\over c^2\partial y^2} \mu_{\rho} \mu_{\sigma}\ ,\label{eq:A6}
\end{equation}
\[ A_{\rho\sigma}= Ag_{\rho\sigma}
  -2{\partial A\over c^2\partial \mu^2} \mu_{\rho} \mu_{\sigma}
  -{\partial A\over c^2\partial y^2} \mu_{\rho} s_{\sigma}
  - {\partial C\over c^2\partial y^2} s_{\rho} s_{\sigma}
  -2{\partial C\over c^2\partial \mu^2} s_{\rho} \mu_{\sigma} \  . \]

The required characteristic equations are to be obtained by substituting
the formulae (\ref{eq:A1}) and (\ref{eq:A2}) in the discontinuities of
the relevant dynamical equations, which are the integrability condition
(\ref{eq:3.1}) for (\ref{eq:2.1}), together with (\ref{eq:2.2}),
(\ref{eq:2.3}) and (\ref{eq:2.4}).  The last of these gives
\begin{equation}
  s^\rho\lambda_{[\rho}\hat\Theta_{\sigma]}=0\ , \label{eq:A7}
\end{equation}
the flux conservation equations (\ref{eq:2.2}) and (\ref{eq:2.3}) give
\begin{equation}
  \lambda_\rho\hat n^\rho=0\ ,\label{eq:A8}
\end{equation}
and
\begin{equation}
  \lambda_\rho\hat s^\rho=0\ ,\label{eq:A9}
\end{equation}
while finally the irrotationality condition (\ref{eq:3.1}) gives
\begin{equation}
  \lambda_{[\rho}\hat\mu_{\sigma]}= 0\ ,\label{eq:A10}
\end{equation}
which is interpretable as meaning that $\hat\mu_\rho$ must be proportional
to $\lambda_\rho$.  The weaker condition (\ref{eq:A7}) gives rise to two
qualitatively different possibilities, the one of interest in relation
to the first and second sound modes under consideration here being that
$\hat\Theta_\rho$ should also be proportional to $\lambda_\rho$.  This
condition is expressible by replacing (\ref{eq:A7}) by the formal
analogue of (\ref{eq:A10}), i.e.
\begin{equation}
  \lambda_{[\rho}\hat\Theta_{\sigma]}=0 \ .\label{eq:A11}
\end{equation}

The alternative possibility is the trivial case of a thermal shearing mode
as characterised by
$s^\rho\lambda_\rho=0$ and $ s^\rho\hat\Theta_\rho= 0$.
In the more general non-superfluid case \cite{[9]} the ordinary particle
constituent could also have a shear type mode, but this second kind of
trivial mode is excluded in the superfluid case by the irrotationality
condition (\ref{eq:A10}).  Leaving aside this trivial case, we now
restrict our attention to the sound type modes characterised by the
restriction of (\ref{eq:A7}) to (\ref{eq:A11}).

To deal with the ensuing system of characteristic equations (\ref{eq:A8}),
(\ref{eq:A9}), (\ref{eq:A10}) and (\ref{eq:A11}), we now put it in a
more explicit form by the use of local ``superfluid frame" Minkowski
coordinates $\{x^\rho\}=\{x^{_0}, x^{_1}, x^{_2},x^{_3}\}$
aligned with the relative flow direction at a particular point under
consideration, so that the independent dynamical variables will be given
there by
\begin{equation}
\{\mu^\rho\}=\mu\{1,\, 0, \, 0, \, 0\}\ ,\qquad
\{s^\rho\}= {s_{_{\rm S}}}\{1,\, v,\, 0,\, 0\}\ . \label{eq:A12}
\end{equation}
There will be no loss of generality in taking the characteristic covector to
have the form
\begin{equation}
  \{\lambda_\rho\}=\{-u,\,\cos\theta,\,\sin\theta,\, 0\}\ ,\label{eq:A13}
\end{equation}
where $\theta$ is the angle between the relative flow direction and the
direction of propagation of the perturbation, and $u$ is the
characteristic velocity whose evaluation is the ultimate objective of
the exercise.   The ensuing set of characteristic equations
can be organised in three subsets. Firstly there is a ``superfluid"
subset obtained from (\ref{eq:A8}) and (\ref{eq:A10}) in the form
\begin{equation}
  u\,\hat n^{_0}-\cos\theta\,\hat n^{_1}-\sin\theta\,\hat n^{_2}=0\
      ,\label{eq:A14}
\end{equation}
\begin{equation}
  \cos\theta\, \hat\mu_{_0}+ u\,\hat\mu_{_1}=0\ ,\qquad
\sin\theta\, \hat\mu_{_1}-\cos\theta\,\hat\mu_{_2}=0\ .\label{eq:A15}
\end{equation}
Next there is a ``thermal'' (or ``normal'') subset obtained from
(\ref{eq:A9}) and (\ref{eq:A11}) in the form
\begin{equation}
 u\,\hat s^{_0}-\cos\theta\,\hat s^{_1}-\sin\theta\,\hat s^{_2}= 0\
   ,\label{eq:A16}
\end{equation}
\begin{equation}
  \cos\theta\, \hat\Theta_{_0}+ u\,\hat\Theta_{_1}=0\ ,\hskip 1.0 cm
\sin\theta\, \hat\Theta_{_1}-\cos\theta\,\hat\Theta_{_2}=0\ .\label{eq:A17}
\end{equation}
Finally there is an orthogonal subset containing the remaining
information from (\ref{eq:A10}) and (\ref{eq:A11}) in the form
\begin{equation}
  \hat\mu_{_3}=0\ ,\qquad \hat\Theta_{_3}=0\ .\label{eq:A18}
\end{equation}
This last set (\ref{eq:A18}) will always decouple and can be solved
separately so as to provide the conclusion that orthogonally to the plane
defined by the relative flow and propagation directions, not just the
momentum discontinuty amplitudes but also those of the currents must
vanish:  $\hat n^{_3}= s^{_3} =0\ $.

In the limit when the temperature $\Theta$ and the entropy magnitude,
$s={\cal O}\{\Theta^3\}$ go to zero, one is left only with the first
subset of characteristic equations (\ref{eq:A14}), (\ref{eq:A15})
corresponding to ordinary sound modes, but when entropy is present it
will evidently be necessary to take account also of the second ``normal"
subset (\ref{eq:A16}), (\ref{eq:A17}).  In general this second set will
be coupled with the first by the cross terms in (\ref{eq:A6}), so that
one will obtain a rather intractable quartic characteristic equation for
the propagation speed $u$.  However it transpires that in the ``cool"
limit with which the present work is concerned, i.e.  to lowest order in
the temperature $\Theta$, the first and second subsets will decouple,
giving a pair of quadratic equations that can easily be solved to give
the respective ``first" and ``second" sound speeds in explicit form.

To see how this comes about, we proceed by evaluating the coefficients
involved using the cool equation of state as specified by substitution
of (\ref{eq:8.7}) in (\ref{eq:8.5}).  The lowest order coefficients can
easily be worked out exactly as
\[ C=3\big(c_{_{\rm P}}^2-v^2\big){\cal Q}\ ,\qquad
   C_{_{00}}=\big(3v^2-c_{_{\rm P}}^2\big) c_{_{\rm P}}^2{\cal Q}\ ,\]
\begin{equation}
  C_{_{11}}=\big(3c_{_{\rm P}}^2-v^2\big){\cal Q}\ ,\qquad
   C_{_{01}}=-2v c_{_{\rm P}}^2{\cal Q}\ ,\label{eq:A19}
\end{equation}
in terms of a factor
\begin{equation}
  {\cal Q}={4\tilde\hbar\over 9}\big(c_{_{\rm P}}^2-v^2\big)^{-4/3}
  c_{_{\rm P}}^{-1/3}{s_{_{\rm S}}}^{-2/3} \ ,\label{eq:A20}
\end{equation}
which is unbounded, growing proportionally to $\Theta^{-2}$, in the limit
as $\Theta\rightarrow 0$.  The coefficients of next order are bounded in
this limit, but are functionally more complicated. Nevertheless, to the
order of accuracy that is needed they are simply expressible in terms of
the dilation coefficient $\Phi^2$ defined by (\ref{eq:3.10}) as
\begin{equation}
 B=\Phi^2 ,\qquad
 B_{_{00}}=-{c^4\over c_{_{\rm P}}^2}\,\Phi^2+{\cal O}\{\Theta^4\}\ ,\qquad
 B_{_{11}}=\Phi^2+{\cal O}\{\Theta^4\}\ .\label{eq:A21}
\end{equation}
To the required order of accuracy, the dilation coefficient itself
(which effectively controles the dynamics of the zero temperature limit
 \cite{[13]}) will be given simply by
\begin{equation}
  \Phi^2={n\over \mu}+{\cal O}\{\Theta^4\} \ .\label{eq:A22}
\end{equation}
Those of the remaining coefficients that are not exactly zero
are not just bounded but are characterised more strictly by
\begin{eqnarray}
  A&=&{\cal O}\{\Theta\}\ , \qquad A_{_{00}}={\cal O}\{\Theta\}\ , \qquad
    A_{_{11}}= {\cal O}\{\Theta\}\ , \nonumber\\
 A_{_{01}}&=&{\cal O}\{\Theta\}\ ,\qquad A_{_{10}}={\cal O}\{\Theta\}
   \ ,\qquad B_{_{01}}={\cal O}\{\Theta^4\}\ .\label{eq:A23}
\end{eqnarray}
Thus they all tend to zero in the limit $\Theta\rightarrow 0$, and so
will drop out as far as the present calculation is concerned.

Since the combination ${\cal Q}{s_{_{\rm S}}}^{2/3}$ is bounded, it can
be seen from (\ref{eq:A23} that the characteristic equation obtained as
a condition of vanishing determinant for the combined system
(\ref{eq:A14}), (\ref{eq:A15}), (\ref{eq:A16}), (\ref{eq:A17}) will be
expressible to lowest order as a product of a pair of quadratic factors
\begin{equation}
  {\cal F}_{_{\rm I}}=B_{_{00}} u^2 +c^2\big(
    B_{_{11}}\cos^2\theta+B\sin^2\theta\big)\ ,\label{eq:A24}
\end{equation}
\[ {\cal F}_{_{\rm I\!I}}=\Big( C\big( C_{_{11}} u^2 +
   2C_{_{01}}\cos\theta\, u +C_{_{00}} \cos^2\theta\big)+\big( C_{_{00}}
   C_{_{11}} -C_{_{01}}{^{\!  2}}\big) \sin^2\theta\, \Big) {s_{_{\rm
   S}}}^{4/3} \]
in the form
\begin{equation}
 {\cal F}_{_{\rm I}}{\cal F}_{_{\rm I\!I}}={\cal O}\{\Theta^4\}\ ,
\label{eq:A25}
\end{equation}
in which the left hand side is finite while the right hand side tends to
zero as $\Theta\rightarrow 0$.  At lowest order we thus obtain an
effective decoupling into two factors one or other of which must vanish
separately.  The first alternative, ${\cal F}_{_{\rm I}}={\cal
O}\{\Theta^4\}$, will characterise ordinary ``first" sound modes, and is
expressible using the explicit expressions (\ref{eq:A21}) simply as
\begin{equation}
  u^2-c_{_{\rm P}}^2={\cal O}\{\Theta^4\}\ .\label{eq:A26}
\end{equation}
It can thus be seen that the ``first" sound speed will be given,
independently of the propagation angle $\theta$, by
\begin{equation}
  u=\pm c_{_{\rm P}}+{\cal O}\{\Theta^4\}\ ,\label{eq:A27}
\end{equation}
consistently with the original interpretation of $c_{_{\rm P}}$ as the
ordinary sound speed in the zero temperature limit.  The more
interesting alternative, ${\cal F}_{_{\rm I\!I}}={\cal O}\{\Theta^4\}$
will characterise ``second" sound modes, and is expressible using the
formulae (\ref{eq:A19}) as
\begin{equation}
 \big(3c_{_{\rm P}}^2- v^2\big)u^2-4c_{_{\rm P}}^2\cos\theta\, v u
   -c_{_{\rm P}}^4 +c_{_{\rm P}}^2\big(1+2\cos^2\theta\big)v^2={\cal
   O}\{\Theta^4\}\ .\label{eq:A28}
\end{equation}
The ``second" sound speed is thereby found to be given by
\begin{equation}
 u={2c_{_{\rm P}}^2 v\cos\theta \pm c_{_{\rm P}}\sqrt {\big(c_{_{\rm
  P}}^2-v^2\big)\big(3c_{_{\rm P}}^2-(1+2\cos^2\theta)v^2\big)}\over
  3c_{_{\rm P}}^2-v^2}+{\cal O}\{\Theta^4\} \ .\label{eq:A29}
\end{equation}

It is to be remarked that when the two constituents are relatively
at rest, i.e. when $v=0$, this expression for the second sound speed
 reduces just to
\begin{equation}
  u=\pm{\,c_{_{\rm P}}\over\sqrt 3}+{\cal O}\{\Theta^4\} \ ,\label{eq:A30}
\end{equation}
whose form has been well known since the original work of Landau \cite{[2]}
in the Newtonian limit characterised by $c_{_{\rm P}}^2<<c^2$.  What is
new here is the demonstration that this formula can be retained without
any change for a relativistic superfluid in which the sound speed
$c_{_{\rm P}}$ may be comparable with the light speed $c$.


\begin{references}
\bibitem{[1]}L.D.  Landau, E.M.  Lifshitz, {\it Course of Theoretical
Physics {\bf 6}:  Fluid Mechanics}, (Pergamon, Oxford, 1959).
\bibitem{[1b]} S.J. Putterman, {Superfluid Hydrodynamics},
(North-Holland, 1974), p.425.
\bibitem{[2]}L.D.  Landau, E.M.  Lifshitz, {\it Course of Theoretical
Physics {\bf 9}:  Statistical Physics}, (Pergamon, Oxford, 1959).
\bibitem{[3]}I.M.  Khalatnikov, {\it Introduction to the Theory of
Superfluidity}, (Benjamin, New York, 1965).
\bibitem{[4]}B.  Carter, I.M.  Khalatnikov, {\it Phys.  Rev.} {\bf D45},
4536 (1992).
\bibitem{[5]}B. Carter, I.M. Khalatnikov, {\it Ann. Phys.} 219, 243 (1992).
\bibitem{[6]}G.L. Comer, D. Langlois, {\it Class. Quantum Grav.} {\bf 11},
709 (1994).
\bibitem{[7]}B. Carter, {\it Nuclear Phys.} {\bf B412}, 345 (1994).
\bibitem{[8]}B.  Carter, in {\it A Random Walk in Relativity and
Cosmology) (Vadya - Raychaudhuri Festschrift, I.A.G.R.G.  1983)}, ed.
N.  Dadhich, J.  Krishna Rao, J.V.  Narlikar, C.V.  Vishveshwara, pp
48-62 (Wiley Eastern, Bombay, 1985).
\bibitem{[9]}B. Carter, in {\it Relativistic Fluid Dynamics (Noto, 1987)}, ed.
A. Anile,Y. Choquet-Bruhat, pp 1-64 (Lecture Notes in Mathematics
{\bf 1385}, Springer-Verlag, Heidelberg, 1989).
\bibitem{[10]}V.V. Lebedev, I.M. Khalatnikov, {\it Sov. Phys. J.E.T.P.}
{\bf 56}, 923 (1982).
\bibitem{[11]}I.M.  Khalatnikov, V.V.  Lebedev, {\it Physics Letters}
{\bf 91 A}, 70 (1982).
\bibitem{[12]}B.  Carter, I.M.  Khalatnikov, {\it Rev.  Math.  Phys.}
{\bf 6}, 277 (1994).
\bibitem{[13]}B.  Carter, ``Axionic vorticity variational formulation
for relativistic perfect fluids", to be published in {\it Class.
Quantum Grav.} (1994).
\end{references}
\end{document}